\documentclass[twocolumn,aps,preprintnumbers,prl,
,amsfonts,amsmath,amssymb,superscriptaddress,floatfix,10pt]{revtex4-1}

\usepackage[dvips]{graphicx}
\usepackage{bm}
\graphicspath{{_PaperFigs/}}

\usepackage{color}

\definecolor{ngreen}{rgb}{0.2,0.6,0.2}

\definecolor{golden}{rgb}{0.8,0.6,0.1}

\definecolor{purp}{rgb}{0.8,0.1,0.8}

\definecolor{orange}{rgb}{0.9,0.3,0}
\definecolor{mar}{rgb}{0.6,0.1,0.1}

\newcommand{\affA}{%
Department of Applied Physics, School of Engineering, 
The University of Tokyo,\\
7-3-1 Hongo, Bunkyo-ku, Tokyo 113-8656, Japan}

\newcommand{\affB}{%
School of Engineering and Information Technology, The University of New South Wales, \\ Canberra 2600, ACT, Australia}

\newcommand{\affC}{%
Centre for Quantum Computation and Communication Technology, Australian Research Council}

\newcommand{\affD}{%
Department of Electrical and Computer Engineering, National University of Singapore, 4 Engineering Drive 3, Singapore 117583}

\newcommand{\affE}{%
Department of Physics, National University of Singapore, 2 Science Drive 3, Singapore 117551}

\begin{document} 

\title{ Quantum-Limited Mirror-Motion Estimation }

\author{Kohjiro Iwasawa}
\affiliation{\affA}
\author{Kenzo Makino}
\affiliation{\affA}
\author{Hidehiro Yonezawa}
\email{yonezawa@ap.t.u-tokyo.ac.jp}
\affiliation{\affA}
\author{Mankei Tsang}
\affiliation{\affD}
\affiliation{\affE}
\author{Aleksandar Davidovic}
\affiliation{\affB}
\author{Elanor Huntington}
\affiliation{\affB}
\affiliation{\affC}
\author{Akira Furusawa}
\email{akiraf@ap.t.u-tokyo.ac.jp}
\affiliation{\affA}

\begin{abstract}
  We experimentally demonstrate optomechanical motion and force
  measurements near the quantum precision limits set by the quantum
  Cram\'er-Rao bounds (QCRBs). Optical beams in coherent and
  phase-squeezed states are used to measure the motion of a mirror
  under an external stochastic force. Utilizing optical phase tracking
  and quantum smoothing techniques, we achieve position, momentum, and
  force estimation accuracies close to the QCRBs with the coherent
  state, while estimation using squeezed states shows clear quantum
  enhancements beyond the coherent-state bounds.
\end{abstract}

\maketitle

The advance of science and technology demands increasingly precise
measurements of physical quantities. The probabilistic nature of
quantum mechanics represents a fundamental roadblock. Over the last
few decades, the issue of quantum limits to precision measurements has
been a key driver in the development of quantum measurement theory
\cite{braginsky,*wiseman_milburn,glm_science}. With the recent
technological advances in quantum optical, electrical, atomic, and
mechanical systems, quantum limits are now becoming relevant to many
metrological applications, such as gravitational-wave detection
\cite{schnabel}, force sensing \cite{kippenberg,*aspelmeyer},
magnetometry \cite{budker}, clocks \cite{katori}, and biological measurements \cite{Taylor2013}.

It is now recognized that quantum detection and estimation theory
\cite{helstrom} provides the appropriate framework for the definition
and proof of quantum measurement limits. For parameter estimation and
the mean-square error (MSE) criterion, a widely studied quantum limit
is the quantum Cram\'er-Rao bound (QCRB) \cite{helstrom,glm2011}. For
gravitational-wave astronomy and many other sensing applications, the
estimation of time-varying parameters, commonly called waveforms in
the engineering literature, is more relevant. QCRBs for waveform
estimation were recently derived in Refs.~\cite{twc,*tsang_open},
although there has not yet been any comparison of the waveform QCRBs
with experimental results to demonstrate their relevance to current
technology.

Quantum estimation of an optical phase waveform was recently
demonstrated experimentally \cite{wheatley,yonezawa} using an optical
phase tracking method that measures the phase via homodyne detection
with feedback control \cite{wiseman1995,*berry2002,*berry2006,*armen},
followed by smoothing of the data
\cite{tsl2008,*tsl2009,smooth,*smooth_pra1,*smooth_pra2}. These
experiments demonstrate improvements over heterodyne measurements,
causal filtering \cite{wheatley}, and coherent-state optical beams
when squeezed light is used \cite{yonezawa}, but no comparison with
the QCRBs was made to test the optimality of the experimental
techniques.

In this Letter, we report an experiment that applies the tracking and
smoothing techniques to optomechanical motion sensing. We use optical
probe beams in coherent and phase-squeezed states to measure the
motion of a mirror under an external stochastic force and then compare
the smoothing errors with the waveform QCRBs.  This is the first time
to our knowledge that experimental results have been compared with the
waveform QCRBs.  Through the comparison, we are able to demonstrate
the near-optimality of our measurement method in the case of coherent
states. The squeezed-state results are further away from the QCRBs but
still show clear enhancements over the coherent-state bounds.  Despite
our focus here on a classical mechanical system, our methods can also
be applied to purely quantum systems
\cite{tsl2008,*tsl2009,*smooth,*smooth_pra1,*smooth_pra2,twc,*tsang_open},
making our methods potentially useful for a wide range of quantum
sensing applications
\cite{glm_science,schnabel,kippenberg,*aspelmeyer,budker,katori,Taylor2013}.

\begin{figure}[htb]
\includegraphics[width=85mm,clip]{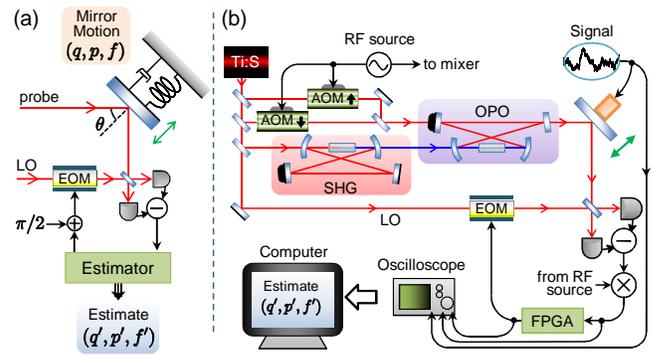}
\caption{(Color online) (a) Schematic of mirror-motion estimation. (b)
  Experimental setup.  LO: Local Oscillator, RF: Radio Frequency,
  Ti:S: Titanium Sapphire laser, AOM: Acousto-Optic Modulator, EOM:
  Electro-Optic Modulator, SHG: Second Harmonic Generator, OPO:
  Optical Parametric Oscillator, FPGA: Field-Programmable Gate Array.
}
\label{fig:setup}
\end{figure}

Figure \ref{fig:setup}(a) shows a schematic of our experiment, where
the mirror motion is approximated as a mass-spring-damper system. The
mirror, driven by a stochastic force, is illuminated by a probe beam
in a coherent state or a phase-squeezed state. The motion of the
mirror shifts the phase of the probe beam. We measure this phase shift
adaptively by homodyne detection (optical phase tracking)
\cite{wiseman1995,*berry2002,*berry2006,*armen,wheatley,yonezawa}, and
estimate the mirror motion from the optical phase measurements
\cite{tsl2008,*tsl2009,*smooth,*smooth_pra1,*smooth_pra2}.

Optical phase tracking allows us to linearize the measurement results
$y(t)$ as
  \begin{align}
    y(t)= \varphi(t) +z(t),
  \end{align}
  where $\varphi(t)$ is the optical phase shift and $z(t)$ is a noise
  term depending on the optical beam statistics
  \cite{wheatley,yonezawa,sup}. The phase shift $\varphi(t)$ of the
  probe beam is caused by the mirror position shift $q(t)$ as
  \begin{align}
   \varphi(t) = ( 2k_0 \cos\theta ) q(t),
  \end{align}
  where $k_0 \cos \theta$ is the wave-vector component parallel to the
  mirror motion and $\theta$ is the reflecting angle as shown in
  Fig.~\ref{fig:setup} (a), fixed at $\pi/4$.  We estimate the mirror
  position $q(t)$, momentum $p(t)$, and external force $f(t)$ from the
  measurement results $y(t)$.  $q(t)$, $p(t)$, $f(t)$, and $z(t)$ are
  assumed to be zero-mean stationary processes.

  Under the linear approximation, the optimal estimate of the mirror
  position is a weighted sum of the measurement results given by
  $q'(t)= \int ^{+\infty}_{-\infty} d\tau J_q(t-\tau) y(\tau)$, where
  $J_q(t)$ is a linear filter and prime indicates an
  estimate. Estimates of momentum $p'(t)$ and external force $f'(t)$
  are similarly defined.  The integration limits are approximated as
  $\pm \infty$ because we use data long before and after $t$ to obtain
  the estimates at the intermediate time $t$ via smoothing
  \cite{tsl2008,*tsl2009,*smooth,*smooth_pra1,*smooth_pra2}.  The
  optimal position filter $J_q(t)$ is obtained by minimizing the MSE
  $\Pi_q=\left \langle [q'(t)-q(t)]^2\right \rangle$, which is
  averaged over the probability measures for $z(t)$ and $q(t)$
  ($\Pi_p$ and $\Pi_f$ are similarly defined). The optimal filters and
  the minimum MSEs are calculated by moving to the frequency domain
  \cite{sup}.  The minimum MSEs $\Pi_x^{\rm min}$ ($x=q,p,f$) are
  given by
  \cite{tsl2008,*tsl2009,*smooth,*smooth_pra1,*smooth_pra2,vantrees,sup}
  \begin{align}
     \Pi_x^{\rm min} &= \int^{+\infty}_{-\infty} \frac{d\omega}{2\pi} 
              \left( 
                  \frac{1}{S_x(\omega)} 
                 +\frac{ | g_{\varphi x}(\omega)|^2 }{S_z(\omega)} 
             \right)^{-1}, 
\label{mmse}  
\end{align}
  where $S_x(\omega)$ ($x=q,p,f,z$) is a spectral density
  defined as $S_x(\omega):=\int^{+\infty}_{-\infty} d\tau \left
    \langle x(t)x(t+\tau) \right \rangle e^{i\omega \tau}$,
  $g_{\varphi x}(\omega)$ is a transfer function that relates the
  optical phase shift $\varphi$ to the target variables ($x=q,p,f$) by
  $\tilde \varphi(\omega)= g_{\varphi x}(\omega) \tilde x(\omega)$,
  with the tilde indicating a Fourier transform.

  We now consider the QCRBs on the MSEs. The waveform QCRBs are
  derived from the quantum properties of the probe beams and prior
  statistics of the target system (mirror motion) and do not depend on
  the measurement and post-processing method.  The QCRBs for our
  situation are \cite{twc,tsang_open}
  \begin{align}
    \Pi_x \geq \int^{+\infty}_{-\infty} 
               \frac{d\omega}{2\pi} 
               \left( \frac{1}{S_x(\omega)} 
                    +|g_{\varphi x}(\omega)|^2 4 S_{\Delta I}(\omega) 
              \right)^{-1},
\label{qcrb}  
\end{align}
where $S_{\Delta I}(\omega)$ is the spectral density of the probe-beam
photon flux.  Comparing Eq.~(\ref{mmse}) with Eq.~(\ref{qcrb}), we
find that $4 S_{\Delta I}(\omega) = 1/S_z(\omega)$ is required for
$\Pi_x^{\rm min}$ to match the QCRBs. This means that, to attain the
QCRBs, (i) the probe beam should be in a minimum-uncertainty state
with respect to the phase and the photon flux, and (ii) the
measurement noise $z(t)$ should consist of intrinsic phase noise only.

Our experiment uses broadband phase-squeezed states, including
coherent states as the small-squeezing limit. The noise term $z(t)$ in
the normalized homodyne outputs can be written in a quadratic
approximation \cite{yonezawa,sup} as
\begin{align}
  \langle z(t)z(\tau)\rangle &= 
                 \frac{\bar R_{\rm sq}}{4|\alpha|^2} \delta(t-\tau), \\
  \bar R_{\rm sq} &= \sigma_\varphi^2 e^{2r_{\rm p}} 
                   + (1-\sigma_\varphi^2) e^{-2r_{\rm m}},
  \label{eq:barRsq}
  \end{align}
  where $r_{\rm m} (r_{\rm p}) $ is the squeezing (anti-squeezing)
  parameter $(r_{\rm p} \geq r_{\rm m} \geq 0)$, $\alpha$ is the
  coherent amplitude of the probe beam, $\sigma_\varphi^2$ is the
  steady-state MSE of the optical phase estimate in the real-time
  feedback loop ($\sigma_\varphi^2 \ll 1$). $\bar{R}_{\rm sq}$ is
  called the {\it effective squeezing factor}~\cite{yonezawa}, which
  takes into account the anti-squeezed amplitude quadrature as well as
  the squeezed phase quadrature.  The noise spectral density
  $S_z(\omega)$ and the photon-flux spectal density $S_{\Delta I}
  (\omega)$ are \cite{sup}
  \begin{align}
   S_z(\omega)&= \frac{\bar R_{\rm sq}}{4|\alpha|^2}, &
   S_{\Delta I}(\omega)&\approx |\alpha|^2 e^{2r_{\rm p}}.
  \end{align}
  Here we assume that the bandwidth of squeezing is broad compared to
  the bandwidth of system parameters, but not too large so
  that the photon-flux fluctuations do not diverge (see Supplemental
  Material \cite{sup}).

  The necessary condition to reach the QCRBs is now given by $e^{2
    r_{\rm p}}=1/\bar R_{\rm sq}$. For coherent states ($r_{\rm
    m}=r_{\rm p}=0$ and $\bar R_{\rm sq}=1$), this condition is always
  satisfied, so QCRB-limited estimation is possible within the
  quadratic approximation.  On the other hand, the squeezed-state QCRB
  is attainable only if (i) the squeezed state is pure ($e^{2 r_{\rm
      p}}= e^{2r_{\rm m}}$) and (ii) the optical phase tracking works
  well enough such that $\sigma_\varphi^2 \simeq 0$. Thus, in a real
  experimental situation, the squeezed-state QCRB is more difficult to
  reach than the coherent-state QCRB. We emphasize however that our
  estimation results are still comparable to the squeeze-state QCRBs
  and better than the coherent-state bounds.

  Figure~\ref{fig:setup}(b) shows our experimental setup.
  A continuous-wave Titanium Sapphire laser is used as a light source
  at 860~nm.  Phase-squeezed states are generated by an optical
  parametric oscillator (OPO) \cite{yonezawa,takeno}.  The OPO is
  driven below threshold by a 430~nm pump beam.  Optical sidebands at
  $\pm$ 5 MHz are used as a carrier beam generated by acousto-optic
  modulators \cite{wheatley,yonezawa}.  To avoid experimental
  complexities, the pump power is fixed at 80 mW, producing squeezing and
  anti-squeezing levels of $-3.62\pm0.26$ dB and 6.00$\pm$0.15 dB. The
  effective squeezing factor, $\bar R_{\rm sq}$, varies from $-3.28$
  dB to $-3.48$ dB depending on the probe amplitude.  To make a
  coherent state, we simply block the pump beam.

  A mirror (12.7~mm in diameter, 1.5~mm in thickness, 0.444~g in
  weight) is attached to a piezoelectric transducer (PZT, weighing
  0.432~g).  We assume the mass of this PZT-mounted mirror to be $m =
  (0.444 + 0.432/3)$~g $= 5.88\times10^{-4}$~kg from the uniformity of
  the PZT \cite{sup}. The transfer function of the PZT-mounted mirror
  (the relation of applied voltage to actual position shift) is
  measured before the estimation experiments.  We use this transfer
  function to construct optimal filters and calculate the QCRBs
  \cite{sup}.

  In the estimation experiments, the PZT-mounted mirror is driven by
  an Ornstein-Uhlenbeck process. This signal is generated by a random
  signal generator followed by a low-pass filter with a cutoff
  frequency of $\lambda = 5.84\times10^4$ rad/s.  We drive the PZT
  within the linear response range so that the external force $f(t)$
  is proportional to the signal. Thus the external force $f(t)$ is
  also an Ornstein-Uhlenbeck process given by
  \begin{align}
   \frac{df(t)}{dt} = -\lambda f(t) +w(t),
   \label{eq:OUpro}
  \end{align}
  where $w(t)$ is a zero-mean white Gaussian noise satisfying $\left
    \langle w(t) w(\tau) \right \rangle =\kappa \delta(t-\tau) $. In
  the experiment, we set $\kappa = 1.67\times10^3$~N$^2$~s$^{-1}$.

  A fraction of the laser beam is used as a local oscillator beam,
  which is optically mixed with the probe beam at a 1:1 beam splitter
  for homodyne detection.  The overall efficiency of the detection is
  87\% \cite{sup}.  The homodyne output is demodulated and recorded
  with an oscilloscope. The measured data are post-processed using a
  computer to produce the estimates.  The demodulated homodyne output
  is also processed by a field programmable gate array (FPGA) for the
  real-time feedback based on Kalman filtering, which approximates the
  mirror motion as a mass-spring-damper system
  \cite{tsl2008,*tsl2009,*smooth,*smooth_pra1,*smooth_pra2}. Note that
  we use this approximate model only for the real-time feedback, not
  for the estimation.  In the experiment, we have another low-gain,
  low-frequency feedback loop to prevent environmental phase drift.

\begin{figure}[!t]
\includegraphics[width=85mm,clip]{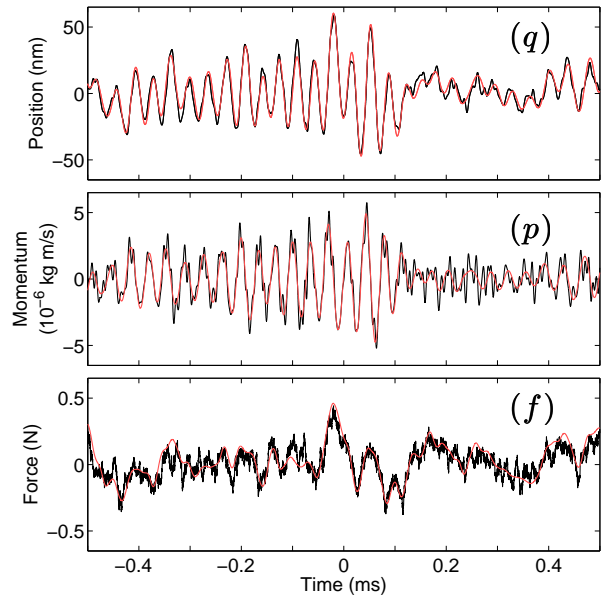}
\caption{ (Color online) Time-domain results for ($q$) position, ($p$)
  momentum, and ($f$) external force, respectively, with $|\alpha|^2=
  6.24\times10^6$ s$^{-1}$ and the probe beam in a phase-squeezed
  state. The black lines are the signals to be estimated. The red
  lines (gray lines in print) are the estimates.  }\label{fig:tplot}
\end{figure}
Figure~\ref{fig:tplot} shows one of the time-domain results for the
mirror-motion estimation with phase-squeezed states. The black lines
are the signals to be estimated (for the evaluation, see Supplemental
Material \cite{sup}).  The external force $f$ is an Ornstein-Uhlenbeck
process given by Eq.~(\ref{eq:OUpro}). The periodic oscillations of
$q$ and $p$ arise from the mechanical resonance of the PZT-mounted
mirror, the frequency of which is $1.76\times10^5$ rad/s
\cite{sup}. The red lines are the estimates, which agree well with the
signals.  This 1~ms long data are obtained with a sampling frequency
of 10 MHz, and are repeated 300 times to evaluate the MSEs.

\begin{figure}[!t]
\includegraphics[width=85mm,clip]{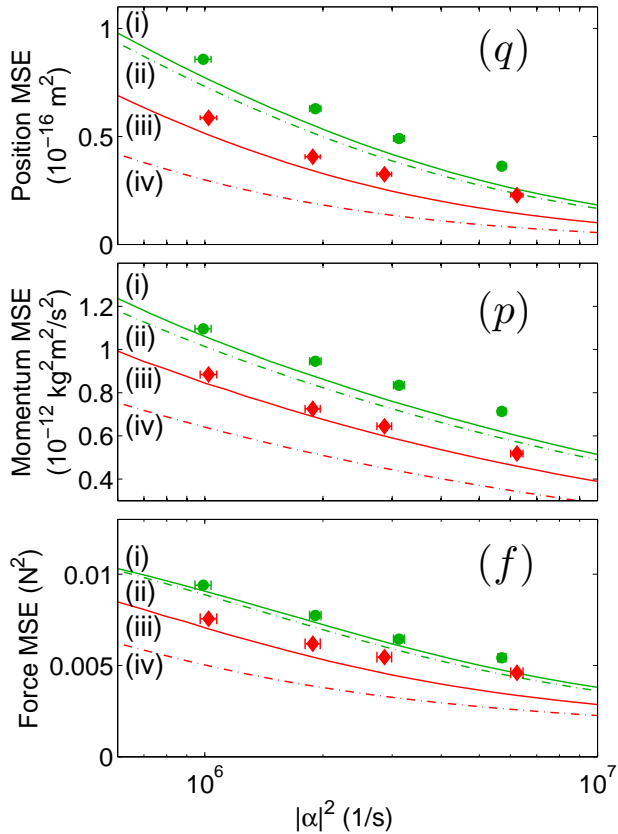}
\caption{ (Color online) Experimental and theoretical MSEs of the
  ($q$) position, ($p$) momentum, and ($f$) external force, plotted
  versus the probe amplitude squared, $|\alpha|^2$. The green circles
  are the results for coherent states, and the red diamonds are those
  for phase-squeezed states.  The green solid curves (traces i) are
  simulated prediction curves of the estimates, which were calculated
  by considering the experimental imperfections. The green dot-dashed
  curves (traces ii) are the coherent-state QCRBs. The red solid lines
  (traces iii) are the simulated prediction curves for a
  phase-squeezed probe beam, where we use the quadratic approximation
  as in Ref.~\cite{yonezawa}. The red dot-dashed curves (trace iv)
  are the squeezed-state QCRBs.  }\label{fig:alpha}
\end{figure}
We perform mirror-motion estimation with probe beams in the coherent
state and the phase-squeezed state, each with four different
amplitudes. Figure~\ref{fig:alpha} shows the $|\alpha|^2$ dependence
of the MSEs of the position, momentum, and external force estimation.
Figure~\ref{fig:alpha} shows three key results.  First key result:
Experimental results agree well with the theoretical predictions
(traces i and iii). The small discrepancies may be attributed to the
low-frequency noise due to environmental phase drift, and slight
changes of the mirror properties (e.g., the resonant frequency) during
the experiment.  Second key result: The experimental results are close
to the waveform QCRBs.  In particular, the experimental results for
coherent states (green circles) are very close to the coherent-state
QCRBs (traces ii). The closeness (i.e., relative differences between
the experimental MSEs and the coherent-state QCRBs) is quantified as
$28 \pm 12 \%$, $15 \pm 6 \%$, and $11 \pm 6 \%$ on average for the
position, momentum and force estimates, respectively.  The small
differences between the prediction curves (traces i) and the
coherent-state QCRBs (traces ii) are attributed to the imperfect
detection efficiency.  The experimental results of squeezed states
(red diamonds) are also comparable to the squeezed-state QCRBs (traces
iv), although the gaps are larger due to the impurity of the squeezed
states.
Third key result: The experimental results for squeezed states show
clear quantum enhancement, mostly overcoming the coherent-state QCRBs.
The quantum enhancements (i.e., relative reduction of MSEs compared to
the coherent-state QCRBs) are quantified as $15 \pm 8 \%$ and $12 \pm
2 \%$ on average for the position and momentum estimates,
respectively. The force estimate at the highest probe amplitude is
slightly worse than the coherent-state QCRB, which should be due to
the low-frequency noise from the environment.
Note that we still observe quantum enhancement of the force estimation
(except the estimate at the highest probe amplitude), which is
quantified as $12 \pm 2 \%$ on average.
  
In conclusion, we have experimentally demonstrated quantum-limited
mirror-motion estimation via optical phase tracking.  Our experiment
reveals that the coherent-state QCRB is almost attainable by our
experimental method. Although the squeezed-state QCRB turns out to be
more difficult to reach because of the impurity of the squeezed
states, quantum enhancement beyond the coherent-state QCRB is clearly
observed. These results demonstrate the potential of our theoretical
and experimental methods for future quantum metrological applications.

\begin{acknowledgments}

  This work was partly supported by PDIS, GIA, G-COE, APSA, FIRST
  commissioned by the MEXT of Japan, SCOPE program of the MIC of
  Japan, the Singapore National Research Foundation under NRF Grant
  No.~NRF-NRFF2011-07, and the Australian Research Council projects
  CE110001029 and DP1094650.  The authors would like to thank Hugo
  Benichi for helpful advice on FPGA digital signal
  processing. H. Y. acknowledges Shuntaro Takeda for constructive
  comments on the manuscript.
  
\end{acknowledgments}

\bibliography{MMEpaper-bib}

\begin{thebibliography}{28}%
\makeatletter
\providecommand \@ifxundefined [1]{%
 \@ifx{#1\undefined}
}%
\providecommand \@ifnum [1]{%
 \ifnum #1\expandafter \@firstoftwo
 \else \expandafter \@secondoftwo
 \fi
}%
\providecommand \@ifx [1]{%
 \ifx #1\expandafter \@firstoftwo
 \else \expandafter \@secondoftwo
 \fi
}%
\providecommand \natexlab [1]{#1}%
\providecommand \enquote  [1]{``#1''}%
\providecommand \bibnamefont  [1]{#1}%
\providecommand \bibfnamefont [1]{#1}%
\providecommand \citenamefont [1]{#1}%
\providecommand \href@noop [0]{\@secondoftwo}%
\providecommand \href [0]{\begingroup \@sanitize@url \@href}%
\providecommand \@href[1]{\@@startlink{#1}\@@href}%
\providecommand \@@href[1]{\endgroup#1\@@endlink}%
\providecommand \@sanitize@url [0]{\catcode `\\12\catcode `\$12\catcode
  `\&12\catcode `\#12\catcode `\^12\catcode `\_12\catcode `\%12\relax}%
\providecommand \@@startlink[1]{}%
\providecommand \@@endlink[0]{}%
\providecommand \url  [0]{\begingroup\@sanitize@url \@url }%
\providecommand \@url [1]{\endgroup\@href {#1}{\urlprefix }}%
\providecommand \urlprefix  [0]{URL }%
\providecommand \Eprint [0]{\href }%
\providecommand \doibase [0]{http://dx.doi.org/}%
\providecommand \selectlanguage [0]{\@gobble}%
\providecommand \bibinfo  [0]{\@secondoftwo}%
\providecommand \bibfield  [0]{\@secondoftwo}%
\providecommand \translation [1]{[#1]}%
\providecommand \BibitemOpen [0]{}%
\providecommand \bibitemStop [0]{}%
\providecommand \bibitemNoStop [0]{.\EOS\space}%
\providecommand \EOS [0]{\spacefactor3000\relax}%
\providecommand \BibitemShut  [1]{\csname bibitem#1\endcsname}%
\let\auto@bib@innerbib\@empty
\bibitem [{\citenamefont {Braginsky}\ and\ \citenamefont
  {Khalili}(1992)}]{braginsky}%
  \BibitemOpen
  \bibfield  {author} {\bibinfo {author} {\bibfnamefont {V.~B.}\ \bibnamefont
  {Braginsky}}\ and\ \bibinfo {author} {\bibfnamefont {F.~Y.}\ \bibnamefont
  {Khalili}},\ }\href@noop {} {\emph {\bibinfo {title} {Quantum Measurement}}}\
  (\bibinfo  {publisher} {Cambridge University Press},\ \bibinfo {address}
  {Cambridge},\ \bibinfo {year} {1992})\BibitemShut {NoStop}%
\bibitem [{\citenamefont {Wiseman}\ and\ \citenamefont
  {Milburn}(2010)}]{wiseman_milburn}%
  \BibitemOpen
  \bibfield  {author} {\bibinfo {author} {\bibfnamefont {H.~M.}\ \bibnamefont
  {Wiseman}}\ and\ \bibinfo {author} {\bibfnamefont {G.~J.}\ \bibnamefont
  {Milburn}},\ }\href@noop {} {\emph {\bibinfo {title} {Quantum Measurement and
  Control}}}\ (\bibinfo  {publisher} {Cambridge University Press},\ \bibinfo
  {address} {Cambridge},\ \bibinfo {year} {2010})\BibitemShut {NoStop}%
\bibitem [{\citenamefont {Giovannetti}\ \emph {et~al.}(2004)\citenamefont
  {Giovannetti}, \citenamefont {Lloyd},\ and\ \citenamefont
  {Maccone}}]{glm_science}%
  \BibitemOpen
  \bibfield  {author} {\bibinfo {author} {\bibfnamefont {V.}~\bibnamefont
  {Giovannetti}}, \bibinfo {author} {\bibfnamefont {S.}~\bibnamefont {Lloyd}},
  \ and\ \bibinfo {author} {\bibfnamefont {L.}~\bibnamefont {Maccone}},\ }\href
  {\doibase 10.1126/science.1104149} {\bibfield  {journal} {\bibinfo  {journal}
  {Science}\ }\textbf {\bibinfo {volume} {306}},\ \bibinfo {pages} {1330}
  (\bibinfo {year} {2004})}\BibitemShut {NoStop}%
\bibitem [{\citenamefont {Schnabel}\ \emph {et~al.}(2010)\citenamefont
  {Schnabel}, \citenamefont {Mavalvala}, \citenamefont {Mc{C}lelland},\ and\
  \citenamefont {Lam}}]{schnabel}%
  \BibitemOpen
  \bibfield  {author} {\bibinfo {author} {\bibfnamefont {R.}~\bibnamefont
  {Schnabel}}, \bibinfo {author} {\bibfnamefont {N.}~\bibnamefont {Mavalvala}},
  \bibinfo {author} {\bibfnamefont {D.~E.}\ \bibnamefont {Mc{C}lelland}}, \
  and\ \bibinfo {author} {\bibfnamefont {P.~K.}\ \bibnamefont {Lam}},\
  }\href@noop {} {\bibfield  {journal} {\bibinfo  {journal} {Nature Commun.}\
  }\textbf {\bibinfo {volume} {1}},\ \bibinfo {pages} {121} (\bibinfo {year}
  {2010})}\BibitemShut {NoStop}%
\bibitem [{\citenamefont {Kippenberg}\ and\ \citenamefont
  {Vahala}(2008)}]{kippenberg}%
  \BibitemOpen
  \bibfield  {author} {\bibinfo {author} {\bibfnamefont {T.~J.}\ \bibnamefont
  {Kippenberg}}\ and\ \bibinfo {author} {\bibfnamefont {K.~J.}\ \bibnamefont
  {Vahala}},\ }\href {\doibase 10.1126/science.1156032} {\bibfield  {journal}
  {\bibinfo  {journal} {Science}\ }\textbf {\bibinfo {volume} {321}},\ \bibinfo
  {pages} {1172} (\bibinfo {year} {2008})}\BibitemShut {NoStop}%
\bibitem [{\citenamefont {Aspelmeyer}\ \emph {et~al.}(2010)\citenamefont
  {Aspelmeyer}, \citenamefont {Gr{\"o}blacher}, \citenamefont {Hammerer},\ and\
  \citenamefont {Kiesel}}]{aspelmeyer}%
  \BibitemOpen
  \bibfield  {author} {\bibinfo {author} {\bibfnamefont {M.}~\bibnamefont
  {Aspelmeyer}}, \bibinfo {author} {\bibfnamefont {S.}~\bibnamefont
  {Gr{\"o}blacher}}, \bibinfo {author} {\bibfnamefont {K.}~\bibnamefont
  {Hammerer}}, \ and\ \bibinfo {author} {\bibfnamefont {N.}~\bibnamefont
  {Kiesel}},\ }\href@noop {} {\bibfield  {journal} {\bibinfo  {journal} {J.\
  Opt.\ Soc.\ Am.\ B}\ }\textbf {\bibinfo {volume} {27}},\ \bibinfo {pages}
  {A189} (\bibinfo {year} {2010})}\BibitemShut {NoStop}%
\bibitem [{\citenamefont {Budker}\ and\ \citenamefont
  {Romalis}(2007)}]{budker}%
  \BibitemOpen
  \bibfield  {author} {\bibinfo {author} {\bibfnamefont {D.}~\bibnamefont
  {Budker}}\ and\ \bibinfo {author} {\bibfnamefont {M.}~\bibnamefont
  {Romalis}},\ }\href@noop {} {\bibfield  {journal} {\bibinfo  {journal}
  {Nature Phys.}\ }\textbf {\bibinfo {volume} {3}},\ \bibinfo {pages} {227}
  (\bibinfo {year} {2007})}\BibitemShut {NoStop}%
\bibitem [{\citenamefont {Katori}(2011)}]{katori}%
  \BibitemOpen
  \bibfield  {author} {\bibinfo {author} {\bibfnamefont {H.}~\bibnamefont
  {Katori}},\ }\href@noop {} {\bibfield  {journal} {\bibinfo  {journal} {Nature
  Photonics}\ }\textbf {\bibinfo {volume} {5}},\ \bibinfo {pages} {203}
  (\bibinfo {year} {2011})}\BibitemShut {NoStop}%
\bibitem [{\citenamefont {Taylor}\ \emph {et~al.}(2013)\citenamefont {Taylor},
  \citenamefont {Janousek}, \citenamefont {Daria}, \citenamefont {Knittel},
  \citenamefont {Hage}, \citenamefont {Bachor},\ and\ \citenamefont
  {Bowen}}]{Taylor2013}%
  \BibitemOpen
  \bibfield  {author} {\bibinfo {author} {\bibfnamefont {M.~A.}\ \bibnamefont
  {Taylor}}, \bibinfo {author} {\bibfnamefont {J.}~\bibnamefont {Janousek}},
  \bibinfo {author} {\bibfnamefont {V.}~\bibnamefont {Daria}}, \bibinfo
  {author} {\bibfnamefont {J.}~\bibnamefont {Knittel}}, \bibinfo {author}
  {\bibfnamefont {B.}~\bibnamefont {Hage}}, \bibinfo {author} {\bibfnamefont
  {H.-A.}\ \bibnamefont {Bachor}}, \ and\ \bibinfo {author} {\bibfnamefont
  {W.~P.}\ \bibnamefont {Bowen}},\ }\href@noop {} {\bibfield  {journal}
  {\bibinfo  {journal} {Nature Photonics}\ }\textbf {\bibinfo {volume} {7}},\
  \bibinfo {pages} {229} (\bibinfo {year} {2013})}\BibitemShut {NoStop}%
\bibitem [{\citenamefont {Helstrom}(1976)}]{helstrom}%
  \BibitemOpen
  \bibfield  {author} {\bibinfo {author} {\bibfnamefont {C.~W.}\ \bibnamefont
  {Helstrom}},\ }\href@noop {} {\emph {\bibinfo {title} {Quantum Detection and
  Estimation Theory}}}\ (\bibinfo  {publisher} {Academic Press},\ \bibinfo
  {address} {New York},\ \bibinfo {year} {1976})\BibitemShut {NoStop}%
\bibitem [{\citenamefont {Giovannetti}\ \emph {et~al.}(2011)\citenamefont
  {Giovannetti}, \citenamefont {Lloyd},\ and\ \citenamefont
  {Maccone}}]{glm2011}%
  \BibitemOpen
  \bibfield  {author} {\bibinfo {author} {\bibfnamefont {V.}~\bibnamefont
  {Giovannetti}}, \bibinfo {author} {\bibfnamefont {S.}~\bibnamefont {Lloyd}},
  \ and\ \bibinfo {author} {\bibfnamefont {L.}~\bibnamefont {Maccone}},\
  }\href@noop {} {\bibfield  {journal} {\bibinfo  {journal} {Nature Photon.}\
  }\textbf {\bibinfo {volume} {5}},\ \bibinfo {pages} {222} (\bibinfo {year}
  {2011})}\BibitemShut {NoStop}%
\bibitem [{\citenamefont {Tsang}\ \emph {et~al.}(2011)\citenamefont {Tsang},
  \citenamefont {Wiseman},\ and\ \citenamefont {Caves}}]{twc}%
  \BibitemOpen
  \bibfield  {author} {\bibinfo {author} {\bibfnamefont {M.}~\bibnamefont
  {Tsang}}, \bibinfo {author} {\bibfnamefont {H.~M.}\ \bibnamefont {Wiseman}},
  \ and\ \bibinfo {author} {\bibfnamefont {C.~M.}\ \bibnamefont {Caves}},\
  }\href {\doibase 10.1103/PhysRevLett.106.090401} {\bibfield  {journal}
  {\bibinfo  {journal} {Phys. Rev. Lett.}\ }\textbf {\bibinfo {volume} {106}},\
  \bibinfo {pages} {090401} (\bibinfo {year} {2011})}\BibitemShut {NoStop}%
\bibitem [{\citenamefont {{Tsang}}(2013)}]{tsang_open}%
  \BibitemOpen
  \bibfield  {author} {\bibinfo {author} {\bibfnamefont {M.}~\bibnamefont
  {{Tsang}}},\ }\href@noop {} {\bibfield  {journal} {\bibinfo  {journal} {ArXiv
  e-prints}\ } (\bibinfo {year} {2013})},\ \Eprint
  {http://arxiv.org/abs/1301.5733v3} {arXiv:1301.5733v3 [quant-ph]}
  \BibitemShut {NoStop}%
\bibitem [{\citenamefont {Wheatley}\ \emph {et~al.}(2010)\citenamefont
  {Wheatley}, \citenamefont {Berry}, \citenamefont {Yonezawa}, \citenamefont
  {Nakane}, \citenamefont {Arao}, \citenamefont {Pope}, \citenamefont {Ralph},
  \citenamefont {Wiseman}, \citenamefont {Furusawa},\ and\ \citenamefont
  {Huntington}}]{wheatley}%
  \BibitemOpen
  \bibfield  {author} {\bibinfo {author} {\bibfnamefont {T.~A.}\ \bibnamefont
  {Wheatley}}, \bibinfo {author} {\bibfnamefont {D.~W.}\ \bibnamefont {Berry}},
  \bibinfo {author} {\bibfnamefont {H.}~\bibnamefont {Yonezawa}}, \bibinfo
  {author} {\bibfnamefont {D.}~\bibnamefont {Nakane}}, \bibinfo {author}
  {\bibfnamefont {H.}~\bibnamefont {Arao}}, \bibinfo {author} {\bibfnamefont
  {D.~T.}\ \bibnamefont {Pope}}, \bibinfo {author} {\bibfnamefont {T.~C.}\
  \bibnamefont {Ralph}}, \bibinfo {author} {\bibfnamefont {H.~M.}\ \bibnamefont
  {Wiseman}}, \bibinfo {author} {\bibfnamefont {A.}~\bibnamefont {Furusawa}}, \
  and\ \bibinfo {author} {\bibfnamefont {E.~H.}\ \bibnamefont {Huntington}},\
  }\href {\doibase 10.1103/PhysRevLett.104.093601} {\bibfield  {journal}
  {\bibinfo  {journal} {Phys. Rev. Lett.}\ }\textbf {\bibinfo {volume} {104}},\
  \bibinfo {pages} {093601} (\bibinfo {year} {2010})}\BibitemShut {NoStop}%
\bibitem [{\citenamefont {Yonezawa}\ \emph {et~al.}(2012)\citenamefont
  {Yonezawa}, \citenamefont {Nakane}, \citenamefont {Wheatley}, \citenamefont
  {Iwasawa}, \citenamefont {Takeda}, \citenamefont {Arao}, \citenamefont
  {Ohki}, \citenamefont {Tsumura}, \citenamefont {Berry}, \citenamefont
  {Ralph}, \citenamefont {Wiseman}, \citenamefont {Huntington},\ and\
  \citenamefont {Furusawa}}]{yonezawa}%
  \BibitemOpen
  \bibfield  {author} {\bibinfo {author} {\bibfnamefont {H.}~\bibnamefont
  {Yonezawa}}, \bibinfo {author} {\bibfnamefont {D.}~\bibnamefont {Nakane}},
  \bibinfo {author} {\bibfnamefont {T.~A.}\ \bibnamefont {Wheatley}}, \bibinfo
  {author} {\bibfnamefont {K.}~\bibnamefont {Iwasawa}}, \bibinfo {author}
  {\bibfnamefont {S.}~\bibnamefont {Takeda}}, \bibinfo {author} {\bibfnamefont
  {H.}~\bibnamefont {Arao}}, \bibinfo {author} {\bibfnamefont {K.}~\bibnamefont
  {Ohki}}, \bibinfo {author} {\bibfnamefont {K.}~\bibnamefont {Tsumura}},
  \bibinfo {author} {\bibfnamefont {D.~W.}\ \bibnamefont {Berry}}, \bibinfo
  {author} {\bibfnamefont {T.~C.}\ \bibnamefont {Ralph}}, \bibinfo {author}
  {\bibfnamefont {H.~M.}\ \bibnamefont {Wiseman}}, \bibinfo {author}
  {\bibfnamefont {E.~H.}\ \bibnamefont {Huntington}}, \ and\ \bibinfo {author}
  {\bibfnamefont {A.}~\bibnamefont {Furusawa}},\ }\href {\doibase
  10.1126/science.1225258} {\bibfield  {journal} {\bibinfo  {journal}
  {Science}\ }\textbf {\bibinfo {volume} {337}},\ \bibinfo {pages} {1514}
  (\bibinfo {year} {2012})}\BibitemShut {NoStop}%
\bibitem [{\citenamefont {Wiseman}(1995)}]{wiseman1995}%
  \BibitemOpen
  \bibfield  {author} {\bibinfo {author} {\bibfnamefont {H.~M.}\ \bibnamefont
  {Wiseman}},\ }\href {\doibase 10.1103/PhysRevLett.75.4587} {\bibfield
  {journal} {\bibinfo  {journal} {Phys. Rev. Lett.}\ }\textbf {\bibinfo
  {volume} {75}},\ \bibinfo {pages} {4587} (\bibinfo {year}
  {1995})}\BibitemShut {NoStop}%
\bibitem [{\citenamefont {Berry}\ and\ \citenamefont
  {Wiseman}(2002)}]{berry2002}%
  \BibitemOpen
  \bibfield  {author} {\bibinfo {author} {\bibfnamefont {D.~W.}\ \bibnamefont
  {Berry}}\ and\ \bibinfo {author} {\bibfnamefont {H.~M.}\ \bibnamefont
  {Wiseman}},\ }\href {\doibase 10.1103/PhysRevA.73.063824} {\bibfield
  {journal} {\bibinfo  {journal} {Phys. Rev. A}\ }\textbf {\bibinfo {volume}
  {65}},\ \bibinfo {pages} {043803} (\bibinfo {year} {2002})}\BibitemShut
  {NoStop}%
\bibitem [{\citenamefont {Berry}\ and\ \citenamefont
  {Wiseman}(2006)}]{berry2006}%
  \BibitemOpen
  \bibfield  {author} {\bibinfo {author} {\bibfnamefont {D.~W.}\ \bibnamefont
  {Berry}}\ and\ \bibinfo {author} {\bibfnamefont {H.~M.}\ \bibnamefont
  {Wiseman}},\ }\href {\doibase 10.1103/PhysRevA.73.063824} {\bibfield
  {journal} {\bibinfo  {journal} {Phys. Rev. A}\ }\textbf {\bibinfo {volume}
  {73}},\ \bibinfo {pages} {063824} (\bibinfo {year} {2006})}\BibitemShut
  {NoStop}%
\bibitem [{\citenamefont {Armen}\ \emph {et~al.}(2002)\citenamefont {Armen},
  \citenamefont {Au}, \citenamefont {Stockton}, \citenamefont {Doherty},\ and\
  \citenamefont {Mabuchi}}]{armen}%
  \BibitemOpen
  \bibfield  {author} {\bibinfo {author} {\bibfnamefont {M.~A.}\ \bibnamefont
  {Armen}}, \bibinfo {author} {\bibfnamefont {J.~K.}\ \bibnamefont {Au}},
  \bibinfo {author} {\bibfnamefont {J.~K.}\ \bibnamefont {Stockton}}, \bibinfo
  {author} {\bibfnamefont {A.~C.}\ \bibnamefont {Doherty}}, \ and\ \bibinfo
  {author} {\bibfnamefont {H.}~\bibnamefont {Mabuchi}},\ }\href {\doibase
  10.1103/PhysRevLett.89.133602} {\bibfield  {journal} {\bibinfo  {journal}
  {Phys. Rev. Lett.}\ }\textbf {\bibinfo {volume} {89}},\ \bibinfo {pages}
  {133602} (\bibinfo {year} {2002})}\BibitemShut {NoStop}%
\bibitem [{\citenamefont {Tsang}\ \emph {et~al.}(2008)\citenamefont {Tsang},
  \citenamefont {Shapiro},\ and\ \citenamefont {Lloyd}}]{tsl2008}%
  \BibitemOpen
  \bibfield  {author} {\bibinfo {author} {\bibfnamefont {M.}~\bibnamefont
  {Tsang}}, \bibinfo {author} {\bibfnamefont {J.~H.}\ \bibnamefont {Shapiro}},
  \ and\ \bibinfo {author} {\bibfnamefont {S.}~\bibnamefont {Lloyd}},\ }\href
  {\doibase 10.1103/PhysRevA.78.053820} {\bibfield  {journal} {\bibinfo
  {journal} {Phys. Rev. A}\ }\textbf {\bibinfo {volume} {78}},\ \bibinfo
  {pages} {053820} (\bibinfo {year} {2008})}\BibitemShut {NoStop}%
\bibitem [{\citenamefont {Tsang}\ \emph {et~al.}(2009)\citenamefont {Tsang},
  \citenamefont {Shapiro},\ and\ \citenamefont {Lloyd}}]{tsl2009}%
  \BibitemOpen
  \bibfield  {author} {\bibinfo {author} {\bibfnamefont {M.}~\bibnamefont
  {Tsang}}, \bibinfo {author} {\bibfnamefont {J.~H.}\ \bibnamefont {Shapiro}},
  \ and\ \bibinfo {author} {\bibfnamefont {S.}~\bibnamefont {Lloyd}},\ }\href
  {\doibase 10.1103/PhysRevA.79.053843} {\bibfield  {journal} {\bibinfo
  {journal} {Phys. Rev. A}\ }\textbf {\bibinfo {volume} {79}},\ \bibinfo
  {pages} {053843} (\bibinfo {year} {2009})}\BibitemShut {NoStop}%
\bibitem [{\citenamefont {Tsang}(2009{\natexlab{a}})}]{smooth}%
  \BibitemOpen
  \bibfield  {author} {\bibinfo {author} {\bibfnamefont {M.}~\bibnamefont
  {Tsang}},\ }\href {\doibase 10.1103/PhysRevLett.102.250403} {\bibfield
  {journal} {\bibinfo  {journal} {Phys. Rev. Lett.}\ }\textbf {\bibinfo
  {volume} {102}},\ \bibinfo {pages} {250403} (\bibinfo {year}
  {2009}{\natexlab{a}})}\BibitemShut {NoStop}%
\bibitem [{\citenamefont {Tsang}(2009{\natexlab{b}})}]{smooth_pra1}%
  \BibitemOpen
  \bibfield  {author} {\bibinfo {author} {\bibfnamefont {M.}~\bibnamefont
  {Tsang}},\ }\href@noop {} {\bibfield  {journal} {\bibinfo  {journal} {Phys.
  Rev. A}\ }\textbf {\bibinfo {volume} {80}},\ \bibinfo {pages} {033840}
  (\bibinfo {year} {2009}{\natexlab{b}})}\BibitemShut {NoStop}%
\bibitem [{\citenamefont {Tsang}(2010)}]{smooth_pra2}%
  \BibitemOpen
  \bibfield  {author} {\bibinfo {author} {\bibfnamefont {M.}~\bibnamefont
  {Tsang}},\ }\href {\doibase 10.1103/PhysRevA.81.013824} {\bibfield  {journal}
  {\bibinfo  {journal} {Phys. Rev. A}\ }\textbf {\bibinfo {volume} {81}},\
  \bibinfo {pages} {013824} (\bibinfo {year} {2010})}\BibitemShut {NoStop}%
\bibitem [{sup()}]{sup}%
  \BibitemOpen
  \href@noop {} {}\bibinfo {note} {See Supplementary Material.}\BibitemShut
  {Stop}%
\bibitem [{\citenamefont {Van~Trees}(2001)}]{vantrees}%
  \BibitemOpen
  \bibfield  {author} {\bibinfo {author} {\bibfnamefont {H.~L.}\ \bibnamefont
  {Van~Trees}},\ }\href@noop {} {\emph {\bibinfo {title} {Detection,
  Estimation, and Modulation Theory, Part I.}}}\ (\bibinfo  {publisher} {John
  Wiley \& Sons},\ \bibinfo {address} {New York},\ \bibinfo {year}
  {2001})\BibitemShut {NoStop}%
\bibitem [{\citenamefont {Takeno}\ \emph {et~al.}(2007)\citenamefont {Takeno},
  \citenamefont {Yukawa}, \citenamefont {Yonezawa},\ and\ \citenamefont
  {Furusawa}}]{takeno}%
  \BibitemOpen
  \bibfield  {author} {\bibinfo {author} {\bibfnamefont {Y.}~\bibnamefont
  {Takeno}}, \bibinfo {author} {\bibfnamefont {M.}~\bibnamefont {Yukawa}},
  \bibinfo {author} {\bibfnamefont {H.}~\bibnamefont {Yonezawa}}, \ and\
  \bibinfo {author} {\bibfnamefont {A.}~\bibnamefont {Furusawa}},\ }\href
  {\doibase 10.1364/OE.15.004321} {\bibfield  {journal} {\bibinfo  {journal}
  {Opt. Express}\ }\textbf {\bibinfo {volume} {15}},\ \bibinfo {pages} {4321}
  (\bibinfo {year} {2007})}\BibitemShut {NoStop}%
\bibitem [{\citenamefont {Gardiner}\ and\ \citenamefont
  {Zoller}(2004)}]{GardinerQN}%
  \BibitemOpen
  \bibfield  {author} {\bibinfo {author} {\bibfnamefont {C.}~\bibnamefont
  {Gardiner}}\ and\ \bibinfo {author} {\bibfnamefont {P.}~\bibnamefont
  {Zoller}},\ }\href@noop {} {\emph {\bibinfo {title} {Quantum noise}}},\
  Vol.~\bibinfo {volume} {56}\ (\bibinfo  {publisher} {Springer},\ \bibinfo
  {year} {2004})\BibitemShut {NoStop}%
\end{thebibliography}%

\clearpage

\begin{center}
\large{ Supplemental Material for \\
Quantum-Limited Mirror-Motion Estimation
}
\end{center}

\section{Experimental details}

\begin{figure}[h]
\includegraphics[width=85mm,clip]{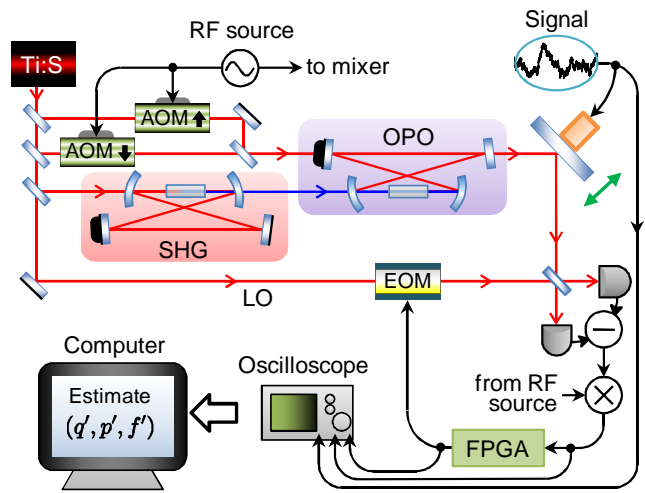}
\caption{
Experimental setup. Ti:S: Titanium Sapphire laser, LO: Local Oscillator, RF: Radio Frequency, AOM: Acousto-Optic Modulator, EOM: Electro-Optic Modulator, SHG: Second Harmonic Generator, OPO: Optical Parametric Oscillator, FPGA: Field Programmable Gate Array.
}\label{fig:expsetup}
\end{figure}

In this section, we will describe the experimental details.  Figure~\ref{fig:expsetup}  shows our experimental setup \cite{yonezawa}. 
A continuous-wave Titanium Sapphire laser was used as a light source at 860 nm. 
Phase squeezed states were generated by an optical parametric oscillator (OPO) of a bow-tie shaped configuration with a periodically polled KTiOPO$_4$ crystal as a nonlinear  optical medium~\cite{takeno}. The OPO was driven below threshold by a 430 nm pump beam, generated by another bow-tie shaped cavity that contains a KNbO$_3$ crystal. The free spectral range and the half width at half maximum of the OPO were 1 GHz and 13 MHz respectively.
Optical sidebands at $\pm$ 5 MHz were used as a carrier beam generated with acousto-optic modulators~\cite{yonezawa, wheatley}.  Note that these optical sidebands are within the OPO's bandwidth. 
 To avoid experimental complexities,  the pump power was fixed to 80 mW giving squeezing and anti-squeezing levels of $-3.62\pm0.26$ dB and 6.00$\pm$0.15 dB respectively.
The effective squeezing factor, $\bar R_{\rm sq}$, varied from $-3.28$ dB to $-3.48$ dB according to the probe amplitude.
Note that $\bar R_{\rm sq}$ takes into account of the anti-squeezing quadratures mixing in the measurement, which cannot be neglected for relatively high squeezing levels. 
It is a trade-off between enhancement from the squeezed quadratures and degradation from the anti-squeezed quadratures, revealing an optimal squeezing level~\cite{yonezawa}.  
The optimal squeezing level differs for each amplitude $|\alpha|$, but the difference is minor for our experimental conditions. 
Since the generated phase squeezed state becomes less robust for higher pumping levels due to the complex locking system, we chose a slightly lower pumping level and did not change it for each $|\alpha|$. 
For comparison to phase squeezed states, we also used coherent states as a probe  by simply blocking the pump beam.

The mirror mounted on a piezoelectric transducer (PZT) was driven by a signal that follows the Ornstein-Uhlenbeck process. 
This  signal was generated with a random signal generator followed by a low-pass filter with a cutoff frequency of $\lambda = 5.84\times10^4$ rad/s. 

A fraction of the laser beam was used as a local oscillator (LO) beam which was passed through a spatial-mode cleaning cavity (not shown in Fig.~\ref{fig:expsetup}) to increase mode matching with the probe beam. 
The probe beam and the LO beam are optically mixed with 1:1 beam splitter for homodyne detection. 
The efficiency of the detection is shown in Table~\ref{tb:imperfection}. 
The homodyne output was demodulated and recorded with an oscilloscope for post processing. 
\begin{table}
\caption{\label{tb:imperfection}  Efficiency of the detection.  }
\begin{ruledtabular}
\begin{tabular}{ll}
Photo diode quantum efficiency & 0.99 \\
Interference efficiency (Visibility) & 0.965~(0.982)\\
Propagation efficiency & 0.981\\
Electrical circuit efficiency (Clearance) & 0.924~(11.2~dB)\\
\hline
Overall efficiency & 0.871 \\
\end{tabular}
\end{ruledtabular}
\end{table}

In the feedback loop, the LO phase is modulated according to the estimated phase. The modulation was performed with a waveguide type electro-optic modulator (EOM). 
The real-time phase estimate used for feedback was processed with a field programmable gate array (FPGA). The delay of our implemented feedback filter was around 400 ns, which is small enough for our current experimental parameters.
Note that we have another low-gain, low-frequency feedback loop to prevent environmental phase drifting.

\section{Modeling the mirror motion}

In this section, we will explain modeling the mirror motion. First, we will consider how to evaluate mass of a PZT-mounted mirror. Then, we will explain transfer function of the PZT-mounted mirror, and the evaluation of {\it true} signals to be estimated. Finally we will describe the mirror motion functions.

\subsection{Mass of a mirror attached to a PZT}
In our experiment, a multilayer PZT (AE0203D04F, NEC/Tokin)  of 3.5 mm$\times$4.5 mm$\times$5.0 mm in size weighing 0.432 g was used. A mirror,  12.7 mm in diameter, 1.5 mm in thickness,  weighing 0.444 g was attached to the PZT with an epoxy-based adhesive. The mass of the mirror attached to the PZT was evaluated as follows.

Let the mass of the PZT and mirror be $M_{\rm p}$ and $M_{\rm m}$, respectively.
Assume that the mass of the PZT is uniform, and that the displacement is proportional at all points,
  \begin{align}
   \Delta l = \frac{l}{L_0}\Delta L.
  \end{align}
Here, the  original  length of the PZT is $L_0$, the overall displacement is $\Delta L$, and the displacement at point $l$ $(0\leq l \leq L_0)$ is $\Delta l$.
Then, the kinetic energy may be calculated as 
  \begin{align}
   E &= \frac{1}{2} M_{\rm m} \left[ \frac{d}{dt}(\Delta L)\right]^2 
       +\int^{L_0}_0  dx \frac{1}{2} \frac{M_{\rm p}}{L_0} 
                                \left[\frac{d}{dt}(\Delta l)\right]^2 
   \nonumber \\
     &= \frac{1}{2} \left( M_{\rm m} + \frac{1}{3}M_{\rm p} \right) 
                                \left[ \frac{d}{dt}(\Delta L)\right]^2. 
  \end{align}
Hence, we assume that $m = M_{\rm m} + M_{\rm p}/3$ $= (0.444 + 0.432/3)$ g $= 5.88\times10^{-4}$ kg. 

\subsection{Transfer function of the PZT-mounted mirror}

Next, we will focus on modeling the transfer function of the PZT-mounted mirror. 
The mass-spring-damper model is referred to as the {\it nominal model}, which is a simplified model that describes the essence of the targeted system. 
On the other hand, a model which best describes the targeted system is referred to as the {\it detailed model}. The detailed model would be the closest measurable model of the targeted system. 
We used this detailed model to construct optimal filters and calculate the QCRBs, while we used the nominal model to realize real-time feedback control.

\begin{figure}[!t]
\includegraphics[width=85mm,clip]{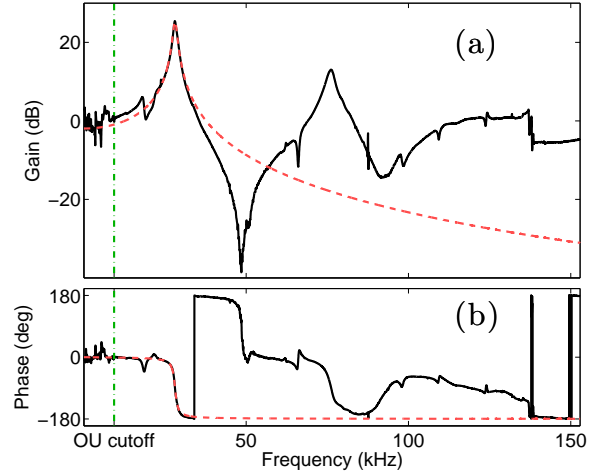}
\caption{
Transfer functions of the PZT-mounted mirror, gain (a) and phase (b).
Black solid lines show the measured transfer function $T_0 (\omega)$ referred to as the detailed model. Red dashed lines show the fitted transfer function of the mass-spring-damper system $T_0^{\rm nom} (\omega)$ referred to as the nominal model. The green dot-dashed line shows the cutoff frequency, $\lambda /2 \pi$, of the Ornstein-Uhlenbeck (OU) process used in the experiment.}
\label{fig:T0}
\end{figure}

We used a Mach-Zehnder interferometer and a network analyzer to measure the transfer function of the PZT-mounted mirror,  $T_0 (\omega)$,  referred to as the detailed model. The black solid lines in Fig.~\ref{fig:T0} show the measured results.
The red dashed lines in Fig.~\ref{fig:T0} show the fitted transfer function of the nominal model $T_0^{\rm nom}(\omega) \propto 1/\left(-m\omega^2+i m \omega \gamma+m\Omega^2 \right)$ where $\gamma$ is the damping coefficient and $\Omega$ is the mechanical resonant frequency. The fitted parameters were $\Omega = 1.76\times10^5$ rad/s and $\gamma=7.66\times10^3$ rad/s. 
  
Note that the external force driving the mirror is generated according to the Ornstein-Uhlenbeck process. 
The cutoff frequency of this process was set to $\lambda = 5.84\times10^4$ rad/s, which is indicated as a green dot-dashed line in Fig.~\ref{fig:T0}. The nominal model is good enough to construct the real-time feedback filter for the experimental conditions. 

\subsection{Evaluation of true signals}

In order to evaluate estimation errors, we need to know the {\it true} position, momentum and external force that are to be estimated (referred to as  the target position, target momentum, and target force).  
We use the full range of the detailed model $T_0(\omega)$  to calculate these target position $q$, momentum $p$, and external force $f$. 
In the mirror motion estimation experiment, we record the voltage $V(t)$ that drives the PZT-mounted mirror. From $V(t)$, $T_0 (\omega)$, and the sensitivity of the photo detector  $G = 6.96\times10^7$ V/m,  we calculate the target position as
  \begin{align}
      q(t) =&  \mathcal{F}^{-1} 
                         \left[ 
                            \frac{T_0(\omega)}{G} \mathcal{F}[V(t)]  
                        \right], 
          \label{eq:qt}
  \end{align}
where $\mathcal{F}~(\mathcal{F}^{-1})$ denotes the (inverse) Fourier transform.
We use this result to calculate the target momentum,
  \begin{align}
   p(t) = m \frac{d}{dt} q(t).
  \label{eq:pt}
  \end{align}
The voltage applied to the PZT-mounted mirror is within the linear response range so that the target force may be calculated as
  \begin{align}
   f(t) = \beta V(t),
  \label{eq:ft}
  \end{align}
where $\beta = 2.04\times10^{-1}$ N/V.

\subsection{Mirror motion functions}

Mirror motion functions $g_{ij}(\omega) \ (i,j=q,p,f,\varphi)$ are defined such as $\tilde q(\omega)=  g_{qf} (\omega) \tilde f(\omega)$ ($i=q$ and $j=f$) where a tilde indicates the Fourier transform. The mirror motion functions are necessary to derive the optimal filters and the QCRBs. 
Note that the definition leads to $ g_{ij}(\omega) g_{jk}(\omega)= g_{ik}(\omega)$ and $\left [  g_{ij}(\omega) \right]^{-1}=  g_{ji}(\omega)$.

From Eqs. (\ref{eq:qt}) and (\ref{eq:ft}), the function $ g_{qf}(\omega)$ is given as,
   \begin{align}
    g_{q f}(\omega)=& \frac{T_0(\omega)}{ G \beta}.
    \label{eq:gqfw}
   \end{align}
As denoted in the main text, the phase shift of the probe beam is proportional to the position shift as $\varphi(t) = (  2k_0 \cos\theta  )  q(t)$. Then, the other relevant mirror motion functions are derived as follows:
   \begin{align}
     g_{\varphi q}(\omega)=& 2k_0 \cos \theta,  \label{eq:gphiqw} \\
     g_{\varphi p}(\omega)=& g_{\varphi q}(\omega) 
                                 g_{q p}(\omega) 
                              =  \frac{ 2k_0 \cos \theta}{i m\omega}, 
                                                   \label{eq:gphipw}  \\
     g_{\varphi f}(\omega)=& g_{\varphi q}(\omega) 
                                  g_{q f}(\omega) 
                                =2k_0 \cos \theta 
                                      \frac{T_0(\omega)}{ G \beta}, 
                                                     \label{eq:gphifw}
   \end{align}
where we use $ g_{q p}(\omega)=1/(i m \omega)$.

\section{Optimal linear filter and least mean square error}

In this section, we derive the optimal linear filters which minimize mean square errors (MSEs) \cite{vantrees}. 
We will explain the position estimate $q'(t)$ and the least position MSE $\Pi_q^{\rm min}$ as an example. The estimates and MSEs for momentum and force can be derived similarly.

First, let's consider the normalized output of the homodyne detection~\cite{berry2002,yonezawa}, 
  \begin{align}
     \eta(t) &=  \sin [ \varphi (t)- \varphi' (t)]  
               + \frac{ v(t)}{ 2|\alpha|} \sqrt{R_{\rm sq}(t)}, \\
     R_{\rm sq}(t) &= \sin^2[\varphi (t) - \varphi' (t)] e^{2r_{\rm p}}
                    + \cos^2 [\varphi (t) - \varphi' (t)] e^{-2r_{\rm m}}. 
     \label{eq:Rsq}
  \end{align}
Here $r_{\rm m}$ $(r_{\rm p}) $ is the squeezing (anti-squeezing) parameter $(r_{\rm p} \geq r_{\rm m} \geq 0)$, $|\alpha|$ is the coherent amplitude of the probe beam, $v(t)$ denotes  white Gaussian noise with a flat spectral density of $1$, and $\varphi' (t)$ is a real-time phase estimate used for the feedback control. 
This homodyne output can also be applied to coherent states by simply putting $R_{\rm sq} = 1$.   
Following the quadratic approximation shown in Ref.~\cite{yonezawa} gives a good approximation of the homodyne output as 
  \begin{align}
    \eta(t) &\simeq \varphi (t) - \varphi' (t)+z(t).
  \label{eq:HDouteta}
  \end{align}
Here, $z(t)$ is a white Gaussian noise as,
  \begin{align}
   \langle z(t)\rangle &= 0, \\
   \langle z(t)z(\tau)\rangle &= 
           \frac{\bar R_{\rm sq}} 
                {4|\alpha|^2} \delta(t-\tau), \label{eq:broadband_z}\\
   \bar R_{\rm sq} &= 
           \sigma_\varphi^2 e^{2r_{\rm p}} 
            + (1-\sigma_\varphi^2) e^{-2r_{\rm m}}
          \label{eq:barRsq}, \\
   \sigma_\varphi^2 &= \langle [\varphi (t)-\varphi' (t)]^2 \rangle.
  \end{align}
$\bar{R}_{\rm sq}$ is called the {\it effective squeezing factor}~\cite{yonezawa}, which takes into account the anti-squeezed amplitude quadrature as well as the squeezed phase quadrature. 

By adding the real-time phase estimate $\varphi' (t)$ (which is measured in the experiment as well as $\eta(t)$) to $\eta(t)$, we obtain the (modified) measurement result $y(t)$,
   \begin{align}
    y(t) = \eta(t)+\varphi'(t) \simeq \varphi(t) +z(t).
   \end{align}
The linear estimate of position, $q'(t)$, is given as a weighted sum of this $y(t)$,
  \begin{align}
    q'(t)= \int ^{+\infty}_{-\infty} d\tau J_q(t-\tau) y(\tau),
  \end{align}
where $J_q(t)$ is a linear position filter.
Fourier transform of the estimate is calculated as,
  \begin{align}
    \tilde q'(\omega)=& \tilde J_q(\omega) \tilde y(\omega)
                     = \tilde J_q(\omega) \left[
                               \tilde \varphi(\omega)+\tilde z(\omega)
                                         \right]
                     \nonumber \\
                     =& \tilde J_q(\omega) \left[
                         g_{\varphi q}(\omega) \tilde q(\omega)
                                +\tilde z(\omega) \right].
  \end{align}
We define a two-time covariance $\Sigma_q(t,t+\tau)$,
  \begin{align}
    \Sigma_q(t,t+\tau):=\left \langle 
                         \left[ q'(t)-q(t)   
                         \right]
                         \left[ q'(t+\tau)-q(t+\tau)
                         \right] 
                     \right \rangle.
  \end{align}
Note that we stick to steady-state so that $\Sigma_q(t,t+\tau)$ is determined by only $\tau$. 
The Fourier transform of $\Sigma_q(t,t+\tau)$ is defined as,
   \begin{align}
    C_q(\omega):=\int^{+\infty}_{-\infty} d\tau\Sigma_q(t,t+\tau) 
                e^{i\omega\tau}.
   \end{align}
MSE of the position estimation, $\Pi_q$, is given as as,
   \begin{align}
    \Pi_q:=\Sigma_q(t,t)=\int^{+\infty}_{-\infty}  
                        \frac{d\omega}{2\pi} C_q(\omega).
   \end{align}
Our aim is to derive the filter $\tilde J_q ( \omega)$ minimizing $\Pi_q$ and obtain the least $\Pi_q$. 

Let's focus on $C_q(\omega)$ because $\Pi_q$ is minimized by minimizing $C_q(\omega)$ at all the $\omega$. 
After some algebra, we find the following:
\begin{align}
    C_q(\omega)=&\left| \tilde J_q(\omega) 
                        g_{\varphi q}(\omega) -1 
               \right|^2 S_q(\omega) 
                +\left| \tilde J_q(\omega)\right|^2 S_z(\omega),
   \end{align}
where $S_k(\omega)$ is a spectral density defined as $S_k(\omega)=\int^{+\infty}_{-\infty} d\tau \left \langle k(t)k(t+\tau) \right \rangle e^{i\omega \tau} \ (k=q,z)$.
By setting $\partial C_q(\omega) / \partial \tilde J_q(\omega)=0$, we obtain the optimal position filter $\tilde J_q^{\rm opt}(\omega)$, 
  \begin{align}
   \tilde J_q^{\rm opt} (\omega)=\frac{
                               g_{\varphi q}^{\ast} (\omega) S_q(\omega)
                                      }
                                      { | g_{\varphi q}(\omega) |^2 
                                    S_q(\omega)+S_z(\omega)
                                      }.
  \end{align}
Accordingly, the least MSE is derived as,
  \begin{align}
     \Pi_q^{\rm min} = \int^{+\infty}_{-\infty} \frac{d\omega}{2\pi} 
            \left( 
                   \frac{1}{S_q(\omega)} 
                  +\frac{| g_{\varphi q}(\omega)|^2}{S_z(\omega)} 
           \right)^{-1}.
  \end{align}
The other optimal filters and MSEs for $p$ and $f$ can be obtained by changing the subscript $q$ to $p$ or $f$.

The spectral densities $S_k(\omega)$ ($k=q,p,f,z$) in our experiment are obtained as follows:
First, $S_z(\omega)$ is easily obtained from Eq. (\ref{eq:broadband_z}),
  \begin{align}
   S_z(\omega)=\frac{\bar R_{\rm sq}} {4|\alpha|^2}.
  \end{align}
The external force $f(t)$ obeys the Ornstein-Uhlenbeck process, 
  \begin{align}
   \frac{df(t)}{dt} &= -\lambda f(t) +w(t), \\
   \left \langle w(t) w(\tau) \right \rangle &= \kappa \delta(t-\tau). 
  \end{align}
Thus, $S_f(\omega)$ is given as,
  \begin{align}
   S_f(\omega)=\frac{\kappa} {\omega^2+\lambda^2}.
  \end{align}
Other spectral densities can be calculated by using the relation $S_i(\omega)=| g_{ij}(\omega)|^2 S_j(\omega)$. From Eq. (\ref{eq:gqfw}) we obtain,
  \begin{align}
   S_q(\omega)&=|g_{qf}(\omega)|^2 S_f(\omega) 
               = \left| \frac{T_0(\omega)}{ G \beta} \right |^2
                 \frac{\kappa} {\omega^2+\lambda^2}, \\
   S_p(\omega)&=|g_{pf}(\omega)|^2 S_f(\omega) 
              = |g_{pq}(\omega)g_{qf}(\omega)|^2 S_f(\omega) 
           \nonumber \\
             &= (m\omega)^2 \left| \frac{T_0(\omega)}{ G \beta} \right |^2
                \frac{\kappa} {\omega^2+\lambda^2}.
  \end{align}
%

\section{Photon flux fluctuation}

In this section, we will derive the spectral density of the photon flux fluctuation $S_{\Delta I}(\omega)$ and discuss the validity of the approximation used in the main text,  $S_{\Delta I}(\omega)\approx |\alpha|^2 e^{2r_{\rm p}}$. 

In order to calculate the photon flux fluctuation, we use an annihilation operator for an electromagnetic field, $a(t)$, which satisfies the commutation relation \cite{GardinerQN},
  \begin{align}
     \left [ a(t), a^\dagger(t') \right]
      &= \delta(t-t'), 
   \label{Eq.a_com 1}\\
     \left [ a(t), a(t') \right]
      &= \left [ a^\dagger(t), a^\dagger(t') \right] = 0. 
  \label{Eq.a_com 2}
  \end{align}
Photon flux $I(t)$ and the mean photon flux $I_0$ are given as,
  \begin{align}
     I(t) =& a^\dagger (t) a(t), \\
     I_0=& \left \langle I(t) \right \rangle.
  \end{align}
We define the Fourier transform of an annihilation operator,
  \begin{align}
   \tilde a(\omega)
   := \int^{+\infty}_{-\infty} dt a(t) e^{i\omega t}.
  \end{align}
Note that this definition leads to $\tilde a^\dagger (\omega)=\tilde a(-\omega)$.
The commutation relation in the frequency domain is derived from Eqs. (\ref{Eq.a_com 1}) and (\ref{Eq.a_com 2}),
  \begin{align}
   \left [
      \tilde a(\omega), \tilde a^\dagger(\omega')
   \right]
      &= 2\pi \delta(\omega-\omega'), 
  \label{Eq.a_com_Freq 1}\\
   \left [
      \tilde a(\omega), \tilde a(\omega')
   \right]
      &=
   \left [
      \tilde a^\dagger(\omega), \tilde a^\dagger(\omega')
   \right]
      = 0. 
  \label{Eq.a_com_Freq 2}   
  \end{align}
Spectral density of $a(t)$ is given as,
  \begin{align}
   R(\omega) :=& \int^{+\infty}_{-\infty} d\tau 
                   \left \langle a^\dagger (t) a(t+\tau) \right \rangle 
                   e^{i\omega \tau}
                 \\ 
              =& \int^{+\infty}_{-\infty} \frac{d\omega_1}{2\pi}  
                    \left \langle 
                        \tilde a^\dagger (\omega_1) \tilde a(\omega) 
                    \right \rangle.
  \end{align}
The mean photon flux $I_0$ is obtained by integrating this spectral density $R(\omega)$,
  \begin{align}
   I_0= \int^{+\infty}_{-\infty} \frac{d\omega}{2\pi}  R(\omega).
  \end{align}

The spectral density of the photon flux fluctuation $S_{\Delta I}(\omega)$ is calculated as,
  \begin{align}
    &S_{\Delta I}(\omega) = \int^{+\infty}_{-\infty} d\tau 
                     \left \langle 
                        \left( I(t) - I_0 
                     \right )
                     \left ( 
                     I(t+\tau) - I_0 
                     \right)
                     \right \rangle e^{i\omega \tau}
                  \nonumber \\
                 &= \int^{+\infty}_{-\infty} 
                             \frac{d\omega_1 d\omega_2 d\omega_3}{(2\pi)^3}
                     \left \langle 
                         \tilde a^\dagger (-\omega_1) 
                         \tilde a^\dagger (-\omega_3) 
                         \tilde a(\omega_2) 
                         \tilde a(\omega-\omega_3) 
                     \right \rangle 
                   \nonumber \\
                 &+ \int^{+\infty}_{-\infty} 
                       \frac{d\omega}{2\pi} R(\omega)
                    -\frac{\delta(\omega)}{2\pi}\int^{+\infty}_{-\infty} 
                       d\omega_1d\omega_2 R(\omega_1)R(\omega_2).
  \label{Eq.PhotonFluxFluc 1}
  \end{align}
To derive  $S_{\Delta I}(\omega)$, we have to calculate the fourth order moment of an annihilation operator. In our case, however, we use a Gaussian state (phase squeezed state), so the second order moment will suffice to describe $S_{\Delta I}(\omega)$. 

Let's assume an annihilation operator of the form, 
  \begin{align}
    \tilde a(\omega) =& 2 \pi \delta(\omega)\left| \alpha \right|
                       + \tilde a_{\rm sq}(\omega),     \label{Eq.a_alp_as} \\
    \tilde a_{\rm sq}(\omega) =& c_{\rm 1a} (\omega) \tilde a_1(\omega)
                         +c_{\rm 1b} (\omega) \tilde a_1^\dagger (-\omega) 
                          \nonumber \\
                       +&c_{\rm 2a} (\omega) \tilde a_2(\omega)
                         +c_{\rm 2b} (\omega) \tilde a_2^\dagger (-\omega), 
   \label{Eq.as}  
  \end{align}
where $|\alpha|$ is a coherent amplitude, $\tilde a_{\rm sq}(\omega)$ represents the squeezing term $(\left\langle \tilde a_{\rm sq}(\omega) \right \rangle=0$), $\tilde a_1(\omega)$ and $\tilde a_2(\omega)$ are vacuum modes. Here we set the amplitude as a real value without loss of generality. The expression of Eq. (\ref{Eq.as}) is valid for any mean-zero Gaussian states including mixed states (i.e., squeezed thermal states), as long as the coefficient $c_{ij}(\omega)$ satisfies the following: 
  \begin{align}
     c_{ij}^\ast (\omega)=& c_{ij} (-\omega), \\
      \left| c_{\rm 1a}(\omega) \right| ^2 
     -\left| c_{\rm 1b}(\omega) \right| ^2 
     +&\left| c_{\rm 2a}(\omega) \right| ^2 
     -\left| c_{\rm 2b}(\omega) \right| ^2 = 1, \\
      c_{\rm 1a}^\ast (\omega) c_{\rm 1b}(\omega)
   +& c_{\rm 2a}^\ast (\omega)c_{\rm 2b}(\omega)
                         \nonumber \\ 
     - c_{\rm 1a} (\omega) c_{\rm 1b}^\ast(\omega)
   -& c_{\rm 2a} (\omega) c_{\rm 2b}^\ast(\omega) =0.
  \end{align}
Here these equations are imposed by the property of the Fourier transform and the commutation relation (Eqs. (\ref{Eq.a_com_Freq 1}) and (\ref{Eq.a_com_Freq 2})).

To describe the photon flux fluctuation of the squeezed states, it is useful to define the quadrature operators,
  \begin{align}
    x_{\rm sq}(\omega) :=& \frac{1}{2} \left[ 
                    \tilde a_{\rm sq}(\omega)+
                    \tilde a_{\rm sq}^\dagger (-\omega)
                               \right],
                  \\
     p_{\rm sq}(\omega) :=& \frac{1}{2i} \left[ 
                    \tilde a_{\rm sq}(\omega)
                   -\tilde a_{\rm sq}^\dagger (-\omega)
                               \right].
  \end{align}
Since we set the amplitude as a real value, $x_{\rm sq}$ ($p_{\rm sq}$) is the anti-squeezing (squeezing) quadrature. 
Photon flux spectrum (except the amplitude contribution), squeezing spectrum and anti-squeezing spectrum (spectral densities of $\tilde a_{\rm sq} (\omega)$, $p_{\rm sq} (\omega)$ and $x_{\rm sq} (\omega)$) are given as,
  \begin{align}
    R_{\rm sq}^{I}(\omega) =& \int^{+\infty}_{-\infty} \frac{d\omega_1}{2\pi} 
                    \left \langle 
                        \tilde a_{\rm sq}^\dagger (\omega_1) 
                        \tilde a_{\rm sq}(\omega) 
                    \right \rangle, \\
    R_{\rm sq}^{-}(\omega) =& \int^{+\infty}_{-\infty} \frac{d\omega_1}{2\pi} 
                    \left \langle 
                        p_{\rm sq}^\dagger (\omega_1) p_{\rm sq}(\omega) 
                    \right \rangle, \\
    R_{\rm sq}^{+}(\omega) =& \int^{+\infty}_{-\infty} \frac{d\omega_1}{2\pi} 
                    \left \langle 
                        x_{\rm sq}^\dagger (\omega_1) x_{\rm sq}(\omega) 
                    \right \rangle. \label{Eq.Def:Rsm}
  \end{align}
Here squeezing and anti-squeezing spectrum satisfy an uncertainty principle, 
$R_{\rm sq}^{+}(\omega) R_{\rm sq}^{-}(\omega) \ge 1/16$ ($\hbar=1/2$).
From Eqs. (\ref{Eq.a_alp_as}) $\sim$ (\ref{Eq.Def:Rsm}), we obtain,
  \begin{align}
    R_{\rm sq}^{\pm}(\omega) =& 
        \frac{1}{4} 
        \left| c_{\rm 1a}(\omega) \pm c_{\rm 1b}(\omega) \right| ^2 
 \nonumber \\
     +& \frac{1}{4} 
        \left| c_{\rm 2a}(\omega) \pm c_{\rm 2b}(\omega) \right| ^2,
 \\
     R_{\rm sq}^{I}(\omega) 
    =& R_{\rm sq}^{+}(\omega) +R_{\rm sq}^{-}(\omega) -\frac{1}{2}.
   \end{align}
Then, after some algebra, Eq. (\ref{Eq.PhotonFluxFluc 1}) is rewritten as,
  \begin{align}
    S_{\Delta I}(\omega) &= 
                           4|\alpha|^2 R_{\rm sq}^+(\omega) 
  \nonumber \\
                        &+\int^{+\infty}_{-\infty} \frac{d\omega_1}{2\pi}
                         \left [ 
                          2 R_{\rm sq}^{+}(\omega_1 + \omega) 
                            R_{\rm sq}^{+}(\omega_1)
                          -\frac{1}{8}
                         \right] 
  \nonumber \\
                         &+\int^{+\infty}_{-\infty} \frac{d\omega_1}{2\pi}
                          \left [ 
                           2 R_{\rm sq}^{-}(\omega_1 + \omega) 
                             R_{\rm sq}^{-}(\omega_1)
                            -\frac{1}{8}
                          \right].  
   \label{Eq.SdI}
  \end{align}

Next, we will assume that the squeezed state has finite bandwidth, and then verify the approximation used in the main text.

Let's consider the squeezing spectrums $R_{\rm sq}^{\pm}(\omega)$ of the standard form \cite{takeno,yonezawa},
  \begin{align}
   R^{\pm}_{\rm sq}(\omega) &=
      \frac{1}{4}+\left( R_{\rm sq}^{\pm}(0)-\frac{1}{4} \right)
                  \frac{ \left( \Delta \omega_{\pm}\right )^2}
                       {\omega^2+\left( \Delta \omega_{\pm} \right)^2},
  \label{eq:Romega} \\
   \frac{\Delta \omega_{+} }{\Delta \omega_{-}} 
    &=\sqrt{
           \frac{1-4R_{\rm sq}^{-}(0)}{4R_{\rm sq}^{+}(0)-1} 
          }, 
  \end{align}
where $\Delta \omega_{-}$ ($\Delta \omega_{+}$) is the bandwidth of squeezing (anti-squeezing), the second equation ensures that $R_{\rm sq}^{+}(\omega) R_{\rm sq}^{-}(\omega) =1/16$ for all $\omega$ when the squeezed state is pure. 
Here we define the averaged squeezing bandwidth $\Delta \omega_0$ and the squeezing parameters ($r_{\rm m}$, $r_{\rm p}$) at the center frequency ($\omega=0$) as,
  \begin{align}
   \Delta \omega_{0} &:=
             \frac{1}{2} \left( 
                 \Delta \omega_{-}  +  \Delta \omega_{+}
                         \right),  
      \\
   e^{-2r_{\rm m}} &:=4 R^{-}_{\rm sq}(0),   
       \\
   e^{2r_{\rm p}}  &:= 4 R^{+}_{\rm sq}(0).
  \end{align}
In the case of the OPO, $\Delta \omega_0$ corresponds to the half width at half maximum of the OPO. 

By inserting Eq. (\ref{eq:Romega}) to Eq. (\ref{Eq.SdI}), we obtain,
  \begin{align}
     S_{\Delta I}(\omega) &= 
              4|\alpha|^2 R_{\rm sq}^{+} (\omega) +I_{\rm sq}
 \nonumber \\
              +\frac{1}{8} & \left[
                 \frac{ (e^{2r_{\rm p}}-1)^2 (\Delta \omega_+)^3 } 
                      { \omega^2+(2 \Delta \omega_+)^2 }
               + \frac{ (1-e^{-2r_{\rm m}})^2 (\Delta \omega_-)^3 }
                      { \omega^2+(2 \Delta \omega_-)^2 }
                            \right],
   \label{Eq.SdIFiniteBand}
           \\
    I_{\rm sq} &:= \int^{+\infty}_{-\infty} \frac{d\omega}{2\pi} 
                   R_{\rm sq}^{I}(\omega)
 \nonumber \\
           &= \frac{1}{8} \left[
                     (e^{2r_{\rm p}}-1) \Delta \omega_+
                    +(e^{-2r_{\rm m}}-1) \Delta \omega_- \right],
    \label{Eq.Isq}
  \end{align}
where $I_{\rm sq}$ is the mean photon flux of the squeezing ($I_0=|\alpha|^2+I_{\rm sq}$). 

If the averaged squeezing bandwidth $\Delta \omega_0$ is much larger than the system parameters, i.e., $\Delta \omega_0 \gg \Omega,\lambda$ ($\Omega$: the resonant frequency of the mirror, $\lambda$: the cutoff frequency of the external force), we may assume $\omega \ll \Delta \omega_{\pm}$. Note that we implicitly assume that $\Delta \omega_{0} \sim \Delta \omega_{-}\sim\Delta \omega_{+}$, which would be justified in our experimental situation as described later. The photon flux fluctuation $S_{\Delta I}(\omega)$ would be approximated to,
  \begin{align}
     S_{\Delta I}(\omega) &\simeq 
              \left( |\alpha|^2 +\xi I_{\rm sq} \right ) e^{2r_{\rm p}},
 \\
    \xi :=& e^{-2r_{\rm p}} \left( 
          1+\frac{1}{4} 
            \frac{(e^{2r_{\rm p}}-1)^{3/2}+(1-e^{-2r_{\rm m}})^{3/2} }
                 {\sqrt{e^{2r_{\rm p}}-1} - \sqrt{1-e^{-2r_{\rm m}}} }
                  \right),
      \label{Eq.xi_of_xiIsq}
  \end{align}
where the parameter $\xi$ ranges from 1 ($r_{\rm p}=r_{\rm m}=0$) to 1/4 ($r_{\rm p} \rightarrow \infty$). If $\xi I_{\rm sq} \ll |\alpha|^2$, 
  \begin{align}
     S_{\Delta I}(\omega) &\approx |\alpha|^2 e^{2r_{\rm p}}.
  \end{align}

Let's consider whether these conditions ($\Delta \omega_0 \gg \Omega,\lambda$ and $\xi  I_{\rm sq} \ll |\alpha|^2$) are satisfied under our experimental situation.
The experimental parameters are, $e^{2r_{\rm p}}= 3.98$ ($6.00$ dB), $e^{-2r_{\rm m}}$=0.435 ($-3.62$ dB), $\xi =0.61$, $\Delta \omega_{+} / \Delta \omega_{-}=0.435$, $\Omega= 1.76 \times 10^5$ rad/s, and $\lambda = 5.84 \times 10^4$ rad/s. 
The averaged squeezing bandwidth $\Delta \omega_0$, however, is tricky to determine. As in Ref. \cite{yonezawa}, we utilize only finite bandwidth around a sideband frequency of 5 MHz. 
Thus it is not appropriate to define the squeezing bandwidth $\Delta \omega_0$ as an OPO's bandwidth ($\Delta \omega_{\rm OPO} = 8.2 \times 10^7$ rad/s ). 
We should consider an effective squeezing bandwidth $\Delta \omega_0^{\rm eff}$ which is not unnecessarily large, but still satisfies $\Delta \omega_0^{\rm eff} \gg \Omega,\lambda$. 
\begin{figure}[htb]
\includegraphics[width=85mm,clip]{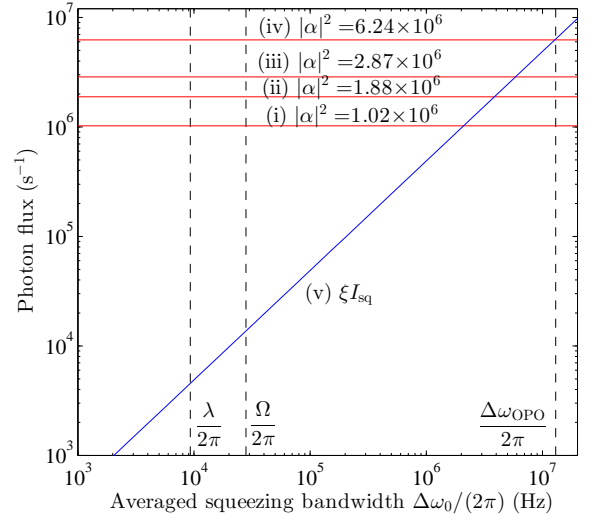}
\caption{
Photon flux versus averaged squeezing bandwidth. Lines (i) to (iv) represent the amplitude squares $|\alpha|^2$ used in the experiment. Trace (v) is the scaled photon flux of squeezing, $\xi I_{\rm sq}$, which is calculated from Eqs. (\ref{Eq.Isq}) and (\ref{Eq.xi_of_xiIsq}). Dashed lines show the specific frequencies in the experiment, $\lambda$, $\Omega$, $\Delta \omega_{\rm OPO}$. 
}
\label{fig:PhotonFluxIsq}
\end{figure}

Figure~\ref{fig:PhotonFluxIsq} shows $\xi I_{\rm sq}$ as a function of the squeezing bandwidth $\Delta \omega_0$. We also plot experimental amplitude squares $|\alpha|^2=$ 1.02, 1.88, 2.87, 6.24 $\times 10^6$ s$^{-1}$. In Fig.~\ref{fig:PhotonFluxIsq}, there is a certain region which satisfies $\Delta \omega_0 > \Omega, \lambda$ and $\xi  I_{\rm sq} < |\alpha|^2$. For example, let's set the effective squeezing bandwidth as ten times of the resonant frequency, $\Delta \omega_0^{\rm eff}= 10 \Omega$ ($> 10 \lambda$). In this case, we obtain $\xi  I_{\rm sq}=1.37 \times 10^5$ s$^{-1}$ which is still an order smaller than the experimental $|\alpha|^2$. Thus we can assume the effective squeezing bandwidth which simultaneously satisfies $\Delta \omega_{0}^{\rm eff} \gg \Omega, \lambda$ and $\xi  I_{\rm sq} \ll |\alpha|^2$. Accordingly we may conclude that the approximation $S_{\Delta I}(\omega) \approx |\alpha|^2 e^{2r_{\rm p}}$ is valid within our experimental conditions.

\end{document}